\documentstyle[graphicx]{mn}
\begin{document}
\title{The termination shock in a striped pulsar wind}
\author[Y.E.Lyubarsky]{Y.E.Lyubarsky\\
Physics Department, Ben-Gurion University, P.O.B.\ 653, Beer-Sheva
84105, Israel; e-mail: lyub@bgumail.bgu.ac.il}
\date{Received/Accepted}
\maketitle
\begin{abstract}
The origin of radio emission from plerions is considered. Recent
observations suggest that radio emitting electrons are presently
accelerated
rather than having been injected at early stages of the plerion
evolution. The observed flat spectra without a low frequency
cutoff imply an acceleration mechanism that raises the average
particle energy by orders of magnitude but leaves most of the
particles at the energy less than about few hundred MeV. It is
suggested that annihilation of the alternating magnetic field at
the pulsar wind termination shock provides the necessary
mechanism. Toroidal stripes of opposite magnetic polarity are
formed in the wind emanated from an obliquely rotating pulsar
magnetosphere (the striped wind). At the termination shock, the
flow compresses and the magnetic field annihilates by driven
reconnection. Jump conditions are obtained for the shock in a
striped wind. It is shown that postshock MHD parameters of the
flow are the same as if the energy of alternating field has
already been converted into the plasma energy upstream the shock.
Therefore the available estimates of the ratio of the Poynting
flux to the matter energy flux, $\sigma$, should be attributed not
to the total upstream Poynting flux but only to that associated
with the average magnetic field. A simple model for the particle
acceleration in the shocked striped wind is presented.
\end{abstract}


\section{Introduction}
Most of the pulsar spin-down power is carried away by a
relativistic, magnetized wind. The pulsar wind injects this energy
into the surrounding nebula in the form of relativistic
electron-positron pairs and magnetic fields therefore such
nebulae, or plerions, emit synchrotron radiation from the radio to
the gamma-ray band. The most famous and well studied example of
the plerion is the Crab Nebula; the spectrum of this source is
measured from about 10 MHz to dozens TeV.
According to MHD models (Rees \& Gunn 1974; Kennel \& Coroniti
1984; Emmering \& Chevalier 1987; Begelman \& Li 1992), the pulsar
wind terminates in a standing shock at a radius defined by the
condition that the confining pressure balances the momentum flux
of the wind. In the Crab case, the shock radius was estimated to
be $3\times 10^{17}$ cm in excellent agreement with observations;
according to Chandra results (Weisskopf et al.\ 2000), the radius
of the shock in the equatorial plane is $4\times 10^{17}$ cm. The
observed brightness and the spectral index distributions are
generally consistent with the assumption that the relativistic
particles are accelerated at the termination shock and then fill
in the nebula, spending the acquired energy on synchrotron
emission and $pdV$ work.

The generic observational feature of plerions is a flat radio
spectrum; the spectral flux may be presented as a power law
function of the frequency, ${\cal F}_{\nu}\propto \nu^{-\alpha}$,
with the spectral index $\alpha=0-0.3$. At high frequencies the
spectrum steepens and the typical spectral slope in the X-ray band
is $\alpha\ga 1$. The overall spectrum of the Crab may be
described as a broken power law with the spectral breaks around
$10^{13}$ Hz, few$\times 10^{15}$ Hz and around 100 keV. The
synchrotron lifetime of the radio emitting electrons (and
positrons, below by electrons I mean both electrons and positrons)
significantly exceeds the plerion age therefore one can not
exclude a priori that they were injected at the very early stage
of the plerion evolution (Kennel \& Coroniti 1984; Atoyan 1999).
In this case the overall spectrum depends on history of the
nebula. However the spectral break at $10^{13}$ Hz may be simply
accounted for the synchrotron burn off effect assuming that
particles emitting from the radio to the optical bands are
injected more or less homogeneously in time with the single power
law energy distribution. This view is strongly supported by
Gallant \& Tuffs (1999, 2002) who found that  the infra-red
spectral index in the central parts of the Crab is close to that
in the radio, and gradually steepens outward. Recent observations
of wisps in the radio band (Bietenholz \& Kronberg 1992;
Bietenholtz, Frail \& Hester 2001) suggest unambiguously that the
radio emitting electrons are accelerated now in the same region as
the ones responsible for the optical to X-ray emission.

If the radio emitting electrons have been injecting into the Crab
Nebula till the present time, the injection rate of electrons
should be about $10^{40}-10^{41}$ s$^{-1}$. It is interesting that
the observed pulsed optical emission from the Crab pulsar suggests
that about the same amount of electrons is ejected from the pulsar
magnetosphere (Shklovsky 1970) so the pulsar does supply the
necessary amount of particles. The observed spectral slope in the
radio band, $\alpha=0.26$, implies the energy distribution of the
injected electrons of the form $N(E)\propto E^{-1.5}$. In this
case, most of the particles find themselves at the low energy end
of the distribution whereas particles at the upper end of the
distribution dominate the energy density of the plasma. Taking
into account that no sign of a low frequency cut-off is observed
in the Crab spectrum down to about 10 MHz whereas the high
frequency break lies in the ultra-violet band (recall that the
break at about $10^{13}$ Hz is attributed to synchrotron cooling
but not to the injected energy distribution), one concludes that
the above distribution extends from $E_{min}\la 100$ MeV to
$E_{max}\sim 10^6$ MeV. At $E>E_{max}$ the distribution becomes
steeper; the spectral slope in the X-ray band, $\alpha=1.1$,
corresponds, with account for the synchrotron burn off effect, to
$N(E)\propto E^{-2.2}$. The distribution further steepens at
$E\sim 10^9$ MeV as it follows from the gamma-ray spectrum of the
Nebula.
Thus the injected electrons have a very
wide energy distribution, their number density being dominated by
low energy electrons whereas the plasma energy density being
dominated by TeV electrons.

The above considerations place severe limits on the pulsar wind
parameters and possible mechanisms of the particle acceleration at
the termination shock. According to the widespread view, the
pulsars emit Poynting-dominated winds however the electro-magnetic
energy is efficiently transferred to the plasma flow such that the
magnetization parameter $\sigma$, defined as the ratio of the
Poynting flux to the kinetic energy flux, is already very small
when the flow enters the termination shock. The mechanisms of such
an energy transfer still remain unclear (the so called
$\sigma$-problem) however dynamics of the flow in the Nebula
suggests that the magnetic pressure is small just beyond the
termination shock, which implies that the Poynting flux just
upstream the shock is very small (Rees \& Gunn 1974; Kundt \&
Krotscheck 1980; Kennel \& Coroniti 1984; Emmering \& Chevalier
1987; Begelman \& Li 1992). In this case it is kinetic energy of
the upstream flow that converts into the energy of accelerated
particles when the plasma flow is randomized at the shock. Then
the characteristic downstream "temperature" is about the upstream
particle kinetic energy, $T\sim mc^2\Gamma_w$, so the average
particle energy does not vary considerably across the shock. A
high energy tail may be formed in the particle energy distribution
(the particle acceleration at the relativistic shocks is
considered by Hoshino et al. (1992); Gallant \& Arons (1994);
Bednarz \& Ostrowski (1998); Gallant \& Achterberg (1999); Kirk et
al. (2000); Achterberg et al. (2001)) however this tail merges, at
its low energy end, with the quasi-thermal distribution at $E\sim
T\sim mc^2\Gamma_w$.
Therefore if the pulsar spin-down power, $L_{sd}$, is converted
into the kinetic energy of the wind, the available upper limit on
the low-frequency break in the Crab spectrum suggests that the
wind Lorentz-factor, $\Gamma_w$, does not exceed few hundred.

On the other hand, the observed flat spectrum may be formed only
if the energy per electron in the wind is much larger than
$m_ec^2\Gamma_w$. Gallant et al.(2002), modifying the original
idea of Hoshino et al. (1992) and Gallant \& Arons (1994),
suggested that the wind is loaded by ions; in this case the radio
emitting electrons are accelerated by resonant absorption of ion
cyclotron waves collectively emitted at the shock front. The
necessary ion injection rate, $\sim L_{sd}/(m_pc^2\Gamma_w)\sim
10^{39}$ s$^{-1}$, vastly exceeds the fiducial Goldreich-Julian
elementary charge loss rate, $\sim 3\times 10^{34}$ s$^{-1}$.
Although one can not exclude by observations that pulsars emit the
required amount of ions,
the available pulsar models do not assume an ion outflow with the
rate exceeding the Goldreich-Julian charge loss rate (Cheng \&
Ruderman 1980; Arons 1983).

Here I consider an alternative possibility, which does not imply a
radical modification of the basic pulsar model.
In the equatorial belt of the wind from an obliquely rotating
pulsar magnetosphere, the sign of the magnetic field alternates
with the pulsar period forming stripes of opposite magnetic
polarity (Michel 1971, 1982; Coroniti 1990; Bogovalov 1999).
Observations of X-ray tori around pulsars (Weisskopf et al.\ 2000;
Helfand, Gotthelf \& Halpern 2001; Gaensler, Pivovaroff \& Garmire
2001; Pavlov et al. 2001; Gaensler et al. 2002; Lu et al. 2002)
suggest that it is in the equatorial belt where most of the wind
energy is transported; theoretical models (e.g., Bogovalov's
(1999) solution for the oblique split monopole magnetosphere)
support this conclusion. Therefore the fate of the striped wind is
of special interest. In the striped wind, the Poynting flux
converts into the particle energy flux when the oppositely
directed magnetic fields annihilate (Coroniti 1990; Lyubarsky \&
Kirk 2001, henceforth LK; Lyutikov 2002; Kirk \& Skj{\ae}raasen
2003). Until now this dissipation mechanism was considered only in
the unshocked wind. It was found that the flow acceleration in the
course of the energy release dilates the dissipation timescale so
that the wind may enter the termination shock before the
alternating field annihilates completely. In this case driven
annihilation of the magnetic field at the shock may provide the
energy necessary to form the flat particle distribution. On the
other hand, the formed distribution may extend down to low enough
energy because the kinetic energy of the flow in the striped wind
is lower than the total energy. The aim of this research is to
consider properties of the termination shock
in the striped wind. 

It will be shown that the alternating field completely annihilates
at the shock so that the downstream parameters of the flow are the
same as if the field has already annihilated upstream the shock.
Therefore the available limits on the upstream magnetization
parameter, $\sigma$, should be attributed not to the total
Poynting flux but to the Poynting flux associated with the
averaged magnetic field (see also Rees \& Gunn 1974;  Kundt \&
Krotscheck 1980). The upstream flow may be Poynting dominated
provided most of the Poynting flux is transferred by alternating
magnetic field. On the other hand, driven reconnection of the
magnetic field within the shock radically alter the particle
acceleration process. It follows from both analytical and
numerical studies that particles are readily accelerated in the
course of the magnetic field reconnection to form the energy
distribution with the power-law index $\beta\sim 1$ (Romanova \&
Lovelace 1992; Zenitani \& Hoshino 2001; Larrabee, Lovelace \&
Romanova 2002). Electrons with such a distribution emit
synchrotron radiation with a flat spectrum therefore Romanova \&
Lovelace (1992) and Birk, Crusius-W\"atzel \& Lesch (2001)
suggested that reconnection plays a crucial role in flat spectrum
extragalactic radio sources. One can naturally assume that radio
emission of plerions is generated by electrons accelerated in the
course of reconnection of the alternating magnetic field at the
pulsar wind termination shock. Then the steeper high-energy
spectrum may be attributed to the Fermi acceleration of particles
preaccelerated in the reconnection process. A simple model for the
particle acceleration in the shocked striped wind is presented
here. It will be shown that the minimal energy of the power law
energy distribution may be low in this case, even less then the
upstream kinetic energy, whereas the energy density of the plasma
will be dominated by high-energy electrons in agreement with the
observed spectra of plerions.

The article is organized as follows. Jump conditions for a shock
in the striped wind are obtained in Sect.\ 2. Making use of these
jump conditions I demonstrate in sect.\ 3 that the alternating
magnetic field dissipates completely at the pulsar wind
termination shock. In sect.\ 4 I discuss particle acceleration by
driven reconnection. Particle acceleration at the shock in the
striped wind is analyzed in sect.\ 5. The obtained results are
summarized in sect.\ 6.

\section{Jump conditions for the shock in a striped wind}
Let us find jump conditions for a shock in a flow with an
alternating magnetic field. The plasma upstream the shock is
assumed to be cold everywhere with the exception of narrow current
sheets separating stripes with opposite magnetic polarity. The
pressure balance implies that the magnetic field strength in
adjacent stripes differs only by sign but not by the absolute
value; on the other hand the width of stripes with opposite
polarity may not be the same so that the average magnetic field
may be nonzero (Fig. 1). One generally finds jump conditions from
the conservation laws and some prescription for the magnetic flux
passing the shock. In the standard MHD shock, the magnetic flux is
conserved; in the case under consideration one should choose more
general prescription to take into account possible annihilation of
the magnetic field at the shock.

\begin{figure}
\includegraphics[scale=0.4]{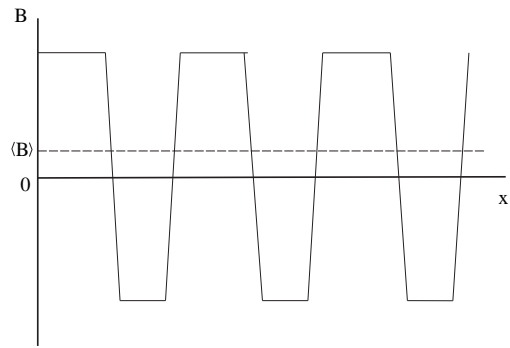}
\caption{The magnetic field in the striped wind. The average field
is shown by dashed line.}
\end{figure}

Outside the shock, the magnetic field is frozen into the plasma,
${\bf E}+(1/c){\bf v\times B}=0$. In the case of interest the
magnetic field lies in the plane of the shock. In the frame where
the flow velocity is perpendicular to the shock plane $E=(v/c)B$.
Then the Faraday law may be written as
$$
\frac{\partial B}{\partial t}+\frac 1c\frac{\partial}{\partial
x}vB=0.
$$
Together with the continuity equation
$$
\frac{\partial}{\partial t}\Gamma n+\frac{\partial}{\partial
x}\Gamma nv=0,
$$
this implies
$$
\frac{B}{n\Gamma}=b,
 \eqno(1)
$$
where $\Gamma$ and $n$ are the Lorentz factor and the proper
plasma density, correspondingly, and $b$ is a constant for each
fluid element. The energy and momentum fluxes are
$$
S=w\Gamma^2v+\frac{EB}{4\pi}c;
$$
$$
F= w\Gamma^2 (v/c)^2+p+\frac{E^2+B^2}{8\pi};
$$
where $p$ and $w$ are the gas pressure and specific enthalpy,
correspondingly.
Taking into account the above considerations, one can write
$$
S={\cal W}v\Gamma^2;\qquad F={\cal W}\Gamma^2(v/c)^2+{\cal
P},\eqno(2)
$$
where the effective pressure and enthalpy are
$$
{\cal P}=p+\frac{b^2n^2}{8\pi},\qquad {\cal
W}=w+\frac{b^2n^2}{4\pi},\eqno(3)
$$
so the flow may be described by the hydrodynamical equations.

In the standard MHD theory, the magnetic flux conserves and the
flow is considered as homogeneous both upstream and downstream the
shock  so that $b$ is a global constant. In our case $b$ is an
alternating function of the lagrangian coordinate however only
$b^2$ enters the conservation laws therefore the only essential
difference from the standard MHD theory is that the modulus $b$
may decrease in the shock because of the field annihilation. Let
us introduce a parameter
$$
\eta=\frac{b_2^2}{b_1^2} \eqno(4)
$$
to measure decreasing in the magnetic flux across the shock. The
indexes 1 and 2 are referred to quantities upstream and downstream
the shock, correspondingly.

One can express the downstream parameters via the upstream ones
making use of the conservation of the particle, energy and
momentum fluxes across the shock. Radiation losses may be safely
neglected at the shock width scale. In the striped wind, one
should take the fluxes averaged over the wave period. Let us
assume, for the sake of simplicity, that the current sheets
between the stripes are so narrow that one can neglect the
contribution of the plasma within the sheets into the conserving
fluxes. Outside the sheets, the magnetic field and plasma density
are constant across a stripe and the adjacent stripes are differ
only by the sign of the magnetic field and by the width. Taking
into account that the energy and momentum fluxes are proportional
to the magnetic field squared, one can write the conservation laws
like in the case of homogeneous magnetic field:
$$
v_1\Gamma_1n_1=v_2\Gamma_2n_2; \eqno(5)
$$
$$
{\cal W}_1v_1\Gamma_1^2={\cal W}_2v_2\Gamma_2^2; \eqno(6)
$$
$$
{\cal W}_1\Gamma_1^2(v_1/c)^2+{\cal P}_1={\cal
W}_2\Gamma_2^2(v_2/c)^2+{\cal P}_2. \eqno(7)
$$
The only formal difference from the case of the standard MHD shock
is in the additional parameter $\eta\le 1$; in the standard shock
$\eta=1$.

One can resolve the above equations as follows. The plasma
downstream the shock is relativistically hot, $w_2=4p_2$,
therefore one can easily eliminate $w$ and $p$ from Eq.(3) and
express ${\cal P}_2$ via ${\cal W}_2$ and $n_2$. Substituting the
obtained relation into Eq.(7), one then eliminates ${\cal W}_2$
making use of Eq.(6) and eliminates $n_2$ making use of Eq.(5).
Taking into account that the plasma upstream the shock is cold,
$p_1=0$ and $w_1=nm_ec^2$, one gets the resulting equation for the
downstream velocity in the form
$$
\left(1+\frac 1{\sigma}\right)
\left(v_1-v_2-\frac{c^2}{4v_2\Gamma_2^2}\right)\frac{\Gamma_1^2v_1}{c^2}+\frac
12
$$$$
= \frac{\eta}4\frac{v_1^2\Gamma_1^2}{v_2^2\Gamma_2^2}-
\frac{v_1\Gamma_1}{4\sigma v_2\Gamma_2},\eqno(8)
$$
where
$$
\sigma=\frac{b_1^2n_1}{4\pi m_ec^2}\eqno(9)
$$
is the magnetization parameter. This equation is exact. In case
$\Gamma_1^2\gg{\rm Max}(\sigma, 1)$ one can take $v_1=c$. Then
Eq.(8) reduces to
$$
(3v_2-c)\left(1+\frac 1{\sigma}\right)=\frac{\eta c}{v_2}(c+v_2).
$$
Now the downstream parameters may be found explicitly:
$$
v_2=\frac c6\left( 1+\chi+\sqrt{1+14\chi+\chi^2}\right); \eqno(10)
$$
$$
n_2=n_1\Gamma_1\frac{\sqrt{2[17-8\chi-\chi^2-(1+\chi)\sqrt{1+14\chi+\chi^2}]}}
{1+\chi+\sqrt{1+14\chi+\chi^2}}; \eqno(11)
$$
$$
\frac{T_2}{m_ec^2}=\frac{\Gamma_1(1+\sigma)}{12\sqrt{2}}
\sqrt{17-8\chi-\chi^2-(1+\chi)\sqrt{1+14\chi+\chi^2}}
$$$$
\times\left(1-\frac{6\chi}{1+\chi+\sqrt{1+14\chi+\chi^2}}\right);
\eqno(12)
$$
where
$$
\chi=\frac{\eta\sigma}{1+\sigma}. \eqno(13)
$$

At $\eta=1$ one recovers the downstream parameters for the
relativistic MHD shock in the homogeneous medium (Kennel \&
Coroniti 1984; Appl \& Camenzind 1988). In this case the quantity
$\chi$ is the ratio of the upstream Poynting flux to the total
energy flux. In a striped wind, $\chi$ is determined by the
Poynting flux corresponding to the magnetic flux passed the shock.
It will be shown in the next section that the alternating magnetic
field annihilates completely at the pulsar wind termination shock;
then $\chi$ is simply the ratio of the Poynting flux associated
with the average magnetic field, $\langle B_1\rangle^2v/4\pi$, to
the total energy flux. At the equator of the flow the average
field is zero, $\alpha=0$, therefore one gets $v_2=c/3$ like at
the nonmagnetized relativistic shock. When the average magnetic
flux is large, $\sigma\gg 1$, $\eta\to 1$, the downstream flow is
relativistic with the Lorentz factor
$$
\Gamma_2=\sqrt{\frac{\sigma}{1+(1-\eta)\sigma}}. \eqno(14)
$$

\section{The termination shock in a striped pulsar wind}
In a striped wind, the magnetic field forms toroidal stripes of
opposite polarity, separated by current sheets (Michel 1971, 1982;
Bogovalov 1999). Such a structure arises in the equatorial belt of
the pulsar wind and may be considered as an entropy wave
propagating through the wind. Usov (1975) and Michel (1982)
noticed that the waves must decay at large distances, since the
current required to sustain them falls off as $r^{-1}$ whereas the
density of available charge carriers in the wind decreases as
$r^{-2}$. Hence the alternating magnetic fields should eventually
annihilate. It was shown in LK that the distance beyond which the
available charge carriers are unable to maintain the necessary
current exceeds the radius of the standing shock where the wind
terminates so that only some fraction of the magnetic energy is
converted into the particle energy before the plasma reaches this
shock front. This fraction depends on the magnetic field
reconnection rate. A lower limit may be obtained assuming that the
dissipation keeps the width of the current sheet to be equal to
the particle Larmor radius, which is roughly the same condition as
that the current velocity is equal to the speed of light
(Corontiti 1990; Michel 1994; LK). Then about 10\% of the Poynting
flux dissipates before the wind reaches the termination shock. At
higher reconnection rate the fraction of the dissipated energy is
larger; assuming that the reconnection velocity is as high as the
sound velocity one can find that the alternating field annihilates
completely before the flow enters the termination shock (Kirk \&
Skj{\ae}raasen 2003). Here I adopt the slow reconnection model by
LK to estimate the flow parameters just upstream the termination
shock.

Neglecting the dependence of the wind parameters on latitude, one
can express them only via the pulsar spin-down power, $L_{sd}$,
and the total amount of the ejected electrons, $\dot N$. LK used
instead the ratio of the gyro-frequency at the light cylinder to
the angular velocity of the neutron star and the so called
multiplicity coefficient arising in the models of the pair
production in pulsar magnetospheres. These parameters may be
expressed via $L_{sd}$ and $\dot N$ taking into account that the
magnetic field at the light cylinder is estimated as
$B_L=\sqrt{L_{sd}/(cR_l)}$, where $R_l=cP/2\pi$ is the light
cylinder radius, $P$ the pulsar period.

The characteristic distance beyond which the available charge
carriers are unable to maintain the necessary current (Eq.(14) in
LK) may be written as
$$
R_{max}=\frac{\pi}2\sqrt{\frac{e^2L_{sd}}{m_e^2c^5}}R_l
, \eqno(15)
$$
where $e$ is the electron charge. 
For the Crab $R_{max}=1.9\times 10^{19}$ cm. Note that the
dimensionless parameter $\hat L$ used by Kirk \& Skj{\ae}raasen
(2003) is proportional to $(R_{max}/R_l)^2$. In the Poynting
dominated flows, dissipation of even a small fraction of the
Poynting flux implies significant acceleration of the flow. While
the dissipated energy is small, the Lorentz factor of the wind
(Eq.(30) in LK) may be estimated as
$$
\Gamma_w=\frac 12\Gamma_{max}\sqrt{\frac R{R_{max}}}, \eqno(16)
$$
where
$$
\Gamma_{max}=\frac{L_{sd}}{m_ec^2\dot N} \eqno(17)
$$
is the Lorentz factor attained by the wind if all the spin-down
power is converted into the kinetic energy of the plasma. Note
that $\Gamma_w/\Gamma_{max}$ is the fraction of the spin-down
power transferred to the plasma therefore the magnetization
parameter of the wind may be written as
$$
\sigma=\frac{\Gamma_{max}}{\Gamma_w}-1. \eqno(18)
$$
 The current sheet width is
conveniently measured by a fraction $\Delta$ of a wavelength $2\pi
R_l$ occupied by the two current sheets; Eq.(31) in LK may be
written as
$$
\Delta=\frac 16\sqrt{\frac R{R_{max}}}.\eqno(19)
$$
In the Crab, the termination shock is observed at the distance
$4\times 10^{17}$ cm (Weisskopf et al.\ 2000) therefore just
upstream the shock $\sigma=13$, $\Delta=0.024$;
$\Gamma_w=4.4\times 10^3/\dot N_{40}$, where $\dot N_{40}=\dot
N/(10^{40}\,{\rm s}^{-1})$.

At the termination shock, the flow sharply decelerates so the
plasma is compressed. The proper density grows $\Gamma_1/\Gamma_2$
times. According to Eqs.(10) and (13), the less $\eta$ (i.e.\ the
larger fraction of the magnetic flux dissipates at the shock) the
less the downstream velocity so one can place a lower limit on the
compression factor assuming $\eta=1$. In this case
$\Gamma_2=\sqrt{\sigma}$; substituting the estimated above
parameters of the Crab wind, one gets
$\Gamma_1/\Gamma_2=\Gamma_w/\sqrt{\sigma}\sim 1000$. Such a huge
compression factor implies significant heating of the plasma,
especially within the current sheets. Let us show that after such
a compression the particle Larmor radius should exceed the
wavelength so that the alternating magnetic field dissipates
completely within the shock.

The downstream temperature, which determines the Larmor radius,
depends on the fraction of the alternating magnetic field
annihilated at the shock. Let us first assume that this fraction
remains small so that the structure sketched in Fig. 1 preserves
downstream the shock. The width of the current sheet in the wind
frame cannot be less than the particle Larmor radius,
$$
\delta=T/eB',\eqno(20)
$$
where $T$ is the plasma temperature within the sheet and $B'$ the
magnetic field in the wind frame of reference. The wavelength in
the wind frame is $2\pi R_l\Gamma$ therefore the minimum fraction
of the wavelength occupied by the two current sheets is
$$
\Delta=T/(\pi R_l\Gamma eB').
$$
The temperature obeys the condition of the hydrostatic equilibrium
of the plasma in the sheet, $n_hT=B'^2/8\pi$, where $n_h$ is the
number density of the plasma in the sheet. The magnetic field is
frozen into the plasma outside the sheet therefore $B'\propto
n_c$, where $n_c$ is the plasma number density outside the sheet.
Now one can write (see Eq.(17) in LK)
$$
\Delta \frac{n_h}{n_c}\propto \frac 1{\Gamma},
$$
The left-hand side of this relation is the fraction of particles
carried in the sheet; of course this fraction should be less than
unity. In the unshocked wind $n_h/n_c=3$ (LK); with the above
estimate for the upstream $\Delta$, one obtains that upstream the
shock this fraction is about 0.1. It was demonstrated above that
$\Gamma$ decreases about 1000 times when the flow passes the
shock, therefore downstream the shock $\Delta n_h/n_c\sim 100$,
which is impossible. This means that contrary to the initial
assumption, the fraction of the magnetic field dissipated within
the shock is not small and $\eta$ defined by Eq.(4) is not close
to unity.

In this case $\chi$ (Eq.(13)) is also not close to unity and then,
according to Eqs.(10) and (12), the downstream flow is non- or
mildly relativistic, $\Gamma_2\sim 1$, whereas the downstream
temperature is about the particle kinetic energy upstream the
shock, $T_2\sim\Gamma_w\sigma m_ec^2$. With the above estimates
for the upstream Lorentz factor and magnetization parameter, one
can easily see that the Larmor radius downstream the shock,
$r_L=T/eB\sim 10^{12}/(B_{-4}\dot N_{40})$ cm, vastly exceeds the
wavelength, $\lambda=2\pi R_l=10^9$ cm, therefore the alternating
magnetic field should completely annihilate at the shock.

This means that the parameter $\chi$ should be determined as the
ratio of the Poynting flux associated with the average magnetic
field, $c\langle B\rangle^2/4\pi$, to the total energy flux.
Therefore the downstream parameters are independent of whether the
alternating magnetic field dissipated upstream the shock or just
at the shock. The available upper limits on the magnetization
parameter in the pulsar wind (Kennel \& Coroniti 1984; Emmering \&
Chevalier 1987; Begelman \& Li 1992) was found from the standard
MHD shock conditions applied to the downstream parameters
estimated from the analysis of the plasma dynamics in the nebula
downstream the shock. In the case of the striped wind these upper
limits should be attributed not to the total Poynting flux but to
the Poynting flux associated with the average magnetic field (see
also Rees \& Gunn 1974; Kundt \& Krotscheck 1980). The upstream
flow may be Poynting dominated provided most of the
electromagnetic energy is carried by the alternating magnetic
field, which annihilates at the termination shock.

\section{Particle acceleration by driven reconnection}
The particle acceleration by driven reconnection may be
qualitatively described as follows. Let us consider a box with two
stripes of the oppositely directed magnetic field and a current
sheet between them (Fig. 2). When the box is compressed, the
electric field ${\bf E}=-(1/c)\bf v\times B$ arises, which has the
same sign in the domains of opposite magnetic polarity. Close
enough to the zero line of the magnetic field, particles may move
freely along this line and gain energy from the electric field. Of
course the real picture should be much more complicated than the
presented one-dimensional sketch. The magnetic reconnection might
proceed in separate X-points but in average the process remains
one-dimensional because the released energy is confined to a layer
around the field reversal.

\begin{figure}
\includegraphics[scale=0.8]{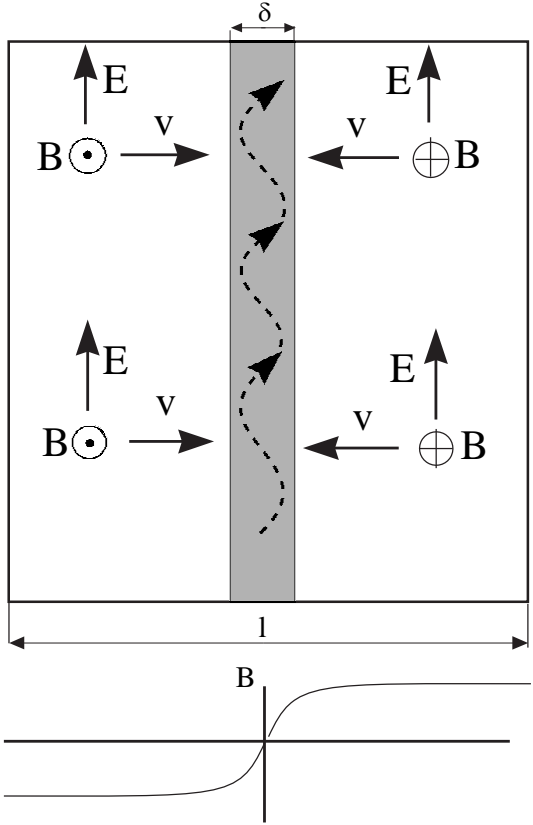}
\caption{A compressing box with oppositely directed magnetic
fields. Distribution of the magnetic field is shown at the bottom.
The current sheet is shaded. The magnetic field is directed
perpendicularly to the plane of the figure. Directions of the
velocities and the electric fields are shown by thick arrows. An
unbounded particle trajectory near the field reversal is shown by
dashed arrows.}
\end{figure}

Acceleration of relativistic electrons close to an X-point was
considered by Romanova \& Lovelace (1992); Zenitani \& Hoshino
(2001), Larrabee et al.(2002) who found a power-law particle
distribution with the slope $\beta\sim 1$. Reconnection in a long
current sheet at a time-scale large enough that particles pass
many X-points has not been considered yet. Let us assume that a
power-law distribution is formed in this case also and estimate
the maximal energy particles attain when the magnetic field
annihilates completely.

Let the particle energy distribution within the current sheet be
$N(\gamma)=K\gamma^{-\beta}$ at $1\le\gamma\le\gamma_m$ with
$\beta\sim 1-2$. At such a distribution function, the particle
density is dominated by low energy electrons,
$$
n_h=\frac K{\beta-1},\eqno(21)
$$
whereas the energy density is dominated by high-energy electrons,
$$
\varepsilon=\frac{m_ec^2K}{2-\beta}\gamma_m^{2-\beta}.\eqno(22)
$$
Let us assume that the power law index $\beta$
remains fixed in the course of compression and only $K$ and
$\gamma_m$ vary. One can find these variations considering
particles and energy balance within the box.

The particle balance is written as
$$
n_h\delta+n_c(l-\delta)=n_{h0}\delta_0+n_{c0}(l_0-\delta_0),\eqno(23)
$$
where $\delta$ is the sheet width, $l$ the box width, $n_h$ and
$n_c$ the particle number densities in the sheet and outside it,
correspondingly, and the index 0 is referred to the initial state.
The energy balance implies that the variation of the total energy
within the box is equal to the work done on the box by the outer
pressure ($=B^2/8\pi$):
$$
d\left[\varepsilon\delta+(l-\delta)\frac{B^2}{8\pi}\right]=-\frac{B^2}{8\pi}dl,
$$
where $\varepsilon$ is the plasma energy density in the sheet.
Taking into account that the magnetic field is frozen into the
cold plasma,
$$
B=bn_c,\eqno(24)
$$
and that the plasma pressure in the sheet, $p=\varepsilon/3$, is
counterbalanced by the magnetic pressure,
$$
\frac{\varepsilon}3=\frac{B^2}{8\pi},\eqno(25)
$$
one can write the energy balance equation in the form
$$
(l+2\delta)\frac{dn_c}{dl}+n_c\frac{d\delta}{dl}+n_c=0.\eqno(26)
$$

Let us assume that the sheet width is equal to the maximal Larmor
radius
$$
\delta=\frac{m_ec^2\gamma_m}{eB}.
$$
The maximal Lorentz factor, $\gamma_m$, is found from Eqs.\ (21),
(22), (24) and (25) as
$$
\gamma_m=\left(\frac{3(2-\beta)}{8\pi(\beta-1)}
\frac{b^2n_c^2}{m_ec^2n_h}\right)^{1/(2-\beta)}.\eqno(27)
$$
Now one can write
$$
\delta=\delta_0\left(\frac{n_c}{n_{c0}}\right)^{\beta/(2-\beta)}
\left(\frac{n_{h0}}{n_{h}}\right)^{1/(2-\beta)}.\eqno(28)
$$
In order to find the plasma parameters in the course of
compression one should solve Eqs.(23), (26) and (28) for
$\delta(l)$, $n_c(l)$, $n_h(l)$ from $l=l_0$ to smaller $l$.  The
magnetic field dissipates completely when the sheet width becomes
comparable with the box width, $\delta\sim l$. At the last stage,
one is unable to separate unambiguously the total plasma volume
onto a hot current sheet and a cold magnetized plasma therefore
the presented equations may be only used to estimate roughly the
final plasma parameters. Taking this into account, one can
simplify the problem even more and solve these equations in the
limit $\delta\ll l$, when they are formally applicable, and take
the limit $\delta=l$ in the obtained solutions.

Let us introduce the dimensionless variable
$\Delta\equiv\delta/l$. In the zeroth order in small $\Delta$,
both Eq.(23) and Eq.(26) reduce to the same equation
$$
n_cl=n_{c0}l_0,\eqno(29)
$$
so the system is nearly degenerate. In order to find the second
equation, one should expand Eqs.(23) and (26) to the first order
in $\Delta$ and eliminate the zeroth order term. Introducing one
more dimensionless variable $Y\equiv n_h/n_c$, one gets
$$
\Delta l\frac{dY}{dl}+(Y-2)l\frac{d\Delta}{dl}+\Delta=0.\eqno(30)
$$
Transforming Eq.(28) to the dimensionless variables $\Delta$ and
$Y$ and substituting $n_c$ from Eq.(29), one obtains
$$
\frac{\Delta}{\Delta_0}=\left(\frac{Y_0l_0}{Yl}\right)^{1/(2-\beta)}.\eqno(31)
$$
Now one can eliminate $l$ from Eqs.(30) and (31) to get finally
$$
(Y-1)\frac{\Delta}Y\frac{dY}{d\Delta}+Y+\beta-4=0.\eqno(32)
$$
The solution to this equation is
$$
\left(\frac{\Delta}{\Delta_0}\right)^{4-\beta}=\frac
{Y_0}Y\left(\frac{Y_0+\beta-4}{Y+\beta-4}\right)^{3-\beta}.
$$
At $\Delta\gg\Delta_0$ (but still $\Delta\ll 1$) $Y$ goes to a
constant, $Y=4-\beta $, independently of the initial conditions.
Taking into account that in the pulsar wind $Y=3$ (LK), one can
take for estimates $Y=Y_0=3$.

Now one can estimate the compression factor, $k\equiv l_0/l$,
necessary for the magnetic field to dissipate completely.
Substituting $\Delta=1$ into Eq.(31), one gets
$$
k=\Delta_0^{\beta-2}.\eqno(33)
$$
Substituting $n_c=kn_{c0}$ into Eq.(27) yields an estimate for the
maximal Lorentz factor attained when the magnetic field dissipates
completely
$$
\gamma_m=\frac 1{\Delta_0}
\left(\frac{(2-\beta)}{2(\beta-1)}\sigma\right)^{1/(2-\beta)},\eqno(34)
$$
where $\sigma=b^2n_{c0}/(4\pi m_ec^2)$ is the initial
magnetization parameter. It follows from Eqs.(16), (18) and (19)
that in the wind upstream the shock $\Delta\sim 1/\sigma$
therefore the particles may be accelerated up to large energies,
$\gamma_m\gg 1$, even at a moderately large $\sigma$.

\section{Particle acceleration at the termination shock in a striped wind}
At the termination shock, the flow is compressed and the energy of
the alternating magnetic field is released. Close to the equator
of the flow, the average magnetic field is small and nearly all
the Poynting flux is transferred to the particles. It was shown in
sect.\ 2 that the downstream velocity is non-relativistic in this
case, therefore the proper density of the plasma increases
$\sim\Gamma_w$ times where $\Gamma_w$ is the wind Lorentz factor
upstream the shock. For typical parameters (see sect.\ 3) this
compression factor significantly exceeds that given by  Eq.\ (33);
this confirms the conclusion that the alternating magnetic field
annihilates completely at the termination shock.

  The flow is decelerated in the shock by the pressure of the hot
downstream plasma and magnetic field. In the collisionless shock
the deceleration scale is about the Larmor radius of those
particles, which make a major contribution to the downstream
pressure. These particles penetrate upstream by their Larmour
radius and exert, via the magnetic field, the decelerating force
on the upstream flow. In the standard MHD shock, the downstream
temperature is about the particle kinetic energy in the upstream
flow therefore the upstream particles penetrate about all the
shock width immediately after they enter the shock. So there is
only one characteristic spatial scale in this case, namely that of
the Larmor radius corresponding to the upstream kinetic energy. In
the shock in the striped wind, the particles gain energy from the
annihilating magnetic field so that the downstream pressure is
determined by particles with the energy significantly exceeding
the upstream kinetic energy. The Larmor radius of these particles
significantly exceeds not only the Larmor radius corresponding to
the kinetic energy in the upstream flow but even the strip width.
Therefore the width of such a shock, or the deceleration scale,
exceeds all "internal" scales in the upstream flow and hence the
the field annihilation proceeds locally in the proper frame like
in the plasma smoothly compressed by an external force. Only when
the alternating field dissipates completely, the particle Larmor
radius, calculated with account of both thermal and kinetic
energy, becomes comparable with the shock width and the flow
decelerates further on like in the standard shock. So one can
roughly separate the shock into two zones. In the first zone, the
flow is decelerated and compressed by the pressure of high energy
particles entering from the second zone. The magnetic field
dissipates in the first zone roughly according to the simple
picture outlined in Sect. 4. The plasma heated by the field
annihilation in the first zone enters the second  one therefore
the second zone resembles the standard shock with a hot upstream
plasma.

The compression factor, $k$, necessary for the magnetic field to
dissipate completely was estimated in the previous section. The
continuity equation in the relativistic flow, $n\Gamma=\it const$,
implies that the flow Lorentz factor at the end of the field
dissipation stage is $\Gamma_d\sim\Gamma_w/k$.  Adopting the
outlined in the previous section picture of the particle
acceleration by driven reconnection, one concludes that a
power-law particle distribution is formed at this stage. In the
proper plasma frame moving with the Lorentz factor $\Gamma_d$,
this distribution extends from $\gamma\sim 1$ up to
$\gamma\sim\gamma_m$.  Now the alternating field is already
dissipated and the flow decelerates further on, from
$\Gamma\sim\Gamma_d$ down to $\Gamma\sim 1$, like in the standard
MHD shock. Applying the particle and energy flux conservation,
$n\Gamma v=\it const$ and $\varepsilon\Gamma^2v=\it const$, one
finds the maximal energy in the particle distribution downstream
the shock, $\gamma_{max}\sim\Gamma_d\gamma_m$. The upper limit on
the minimal energy may be estimated from the condition that the
energy of the most of particles is simply randomized but does not
changes considerably; then $\gamma_{min}\sim\Gamma_d$. Making use
of Eqs.(33) and (34), one gets
$$
\gamma_{min}\sim\Delta^{2-\beta}\Gamma_w;\eqno(35)
$$
$$
\gamma_{max}\sim\frac{\Gamma_w}{\Delta^{\beta-1}}
\left(\frac{2-\beta}{2(\beta-1)}\sigma\right)^{1/(2-\beta)}.\eqno(36)
$$
Substituting $\beta=1.5$ and parameters of the Crab pulsar wind
upstream the termination shock (see sect.3), one gets
$\gamma_{min}\sim 600$; $\gamma_{max}\sim 10^6$, which is roughly
compatible with the parameters inferred from the observed
spectrum.

These simple estimates show that particle acceleration at the
shock in a striped wind may form such a particle distribution that
the energy of most of the particles is significantly less than the
upstream particle kinetic energy whereas the plasma energy density
is dominated by a relatively small amount of high energy
particles. These high energy particles may be accelerated further
on by the 1-st order Fermi mechanism thus forming a high energy
tail at $\gamma>\gamma_{max}$. Recent investigations (Bednarz \&
Ostrowski (1998); Gallant \& Achterberg (1999); Kirk et al.
(2000); Achterberg et al. (2001)) have shown that in
ultra-relativistic shocks, the tail is formed with the power-law
index $\beta=2.2-2.3$ compatible with the observed X-ray spectra
of the Crab and other plerions. So the observed broken power-law
spectrum with a flat low frequency part may be attributed to the
particle acceleration at the termination shock in a striped pulsar
wind.

\section{Conclusions}
The observed spectra of plerions from the radio to the gamma-ray
band imply a very wide electron energy distribution, from less
than few hundreds MeV to $\sim 10^{16}$ eV. Most of electrons are
accumulated at the low-energy end of this distribution. Although
the synchrotron life time of these electrons exceeds the plerion
age, there is strong evidence to suggest that they are accelerated
now together with high energy electrons responsible for the hard
radiation from the nebula (Bietenholz \& Kronberg 1992; Gallant \&
Tuffs 1999, 2002; Bietenholtz et al.\ 2001). The observed flat
radio spectrum implies that the acceleration mechanism transfers
most of the available energy to a small fraction of particles and
retains most of particles at relatively low energy.

It is proposed in this article that the flat energy distribution
is formed in the course of the particle acceleration by driven
reconnection of the alternating magnetic field at the pulsar wind
termination shock. It is widely believed that just upstream the
termination shock the magnetic energy is negligible as compared
with the plasma kinetic energy because plasma dynamics in the
nebula implies small magnetization downstream the shock (Rees \&
Gunn 1974; Kennel \& Coroniti 1984; Emmering \& Chevalier 1987;
Begelman \& Li 1992). However the plasma magnetization and
dynamics depend only on the average upstream magnetic field so the
upstream flow may be Poynting dominated provided most of the
magnetic energy is associated with the alternating magnetic field.
In this case the average particle energy grows significantly at
the shock where the magnetic field annihilates. Therefore the
particle distribution with the power-law index $\beta<2$ is formed
readily.

  The proposed mechanism may also explain why no low-frequency
cutoff is observed in the Crab radio spectrum. At the kinetic
energy dominated shocks, the power-law tail is formed only at the
energies exceeding the downstream temperature, which is about the
particle kinetic energy upstream the shock. Therefore if the Crab
pulsar wind were kinetic energy dominated, only rather low Lorentz
factor of the wind would be compatible with the radio data, which
would imply extremely highly mass loaded wind. In the striped
wind, most of the energy is contained in the magnetic field
therefore the wind Lorentz factor is lower than in the kinetic
energy dominated wind. Moreover the presented here simple model of
the shock in a striped wind predicts that in this case the
downstream temperature may be considerably less than the upstream
particle kinetic energy. Therefore the lack of the low-frequency
turnover in the observed plerion spectra may be naturally
explained within the scope of the proposed model. Of course the
outlined here qualitative picture of the particle acceleration in
plerions may be considered only as preliminary; it may be
justified only by numerical simulations of the shock in a striped
wind.

\section*{Acknowledgments}
I am grateful to David Eichler for stimulating discussions.


\begin{thebibliography}{}
\bibitem{}
Achterberg A., Gallant Y.A., Kirk J.G., Guthmann A.W., 2001,
MNRAS, 328, 393
\bibitem{}
Appl S., Camenzind M.,  1988, A \& A, 206, 258
\bibitem{}
Arons J., 1983, in Positron-electron pairs in astrophysics, Eds.
M.L.Burns, A.K.Harding, R.Ramaty (NY, AIP), p.\ 163
\bibitem{}
Atoyan A.M., 1999, A \& A, 346, L49
\bibitem{}
Bednarz J., Ostrowski M., 1998, Phys.Rev.Lett., 80, 3911
\bibitem{}
Begelman M.C., Li Z.-Y., 1992, ApJ, 397, 187
\bibitem{} Bietenholtz M.F., Frail D.A., Hester J.J., 2001, ApJ,
560, 254
\bibitem{} Bietenholtz M.F., Kronberg P.P., 1992, ApJ, 393, 206
\bibitem{} Birk G.T., Crusius-W\"atzel A.R., Lesch H., 2001, ApJ,
559, 96
\bibitem{}Bogovalov S.V., 1999, A\&A, 349, 101
\bibitem{} Cheng A.F., Ruderman M.A., 1980, ApJ, 235, 576
\bibitem{} Coroniti F.V., 1990, ApJ, 349, 538
\bibitem{} Emmering R.T., Chevalier R.A., 1987, ApJ, 321, 334
\bibitem{} Gaensler B.M., Arons J., Kaspi V.M., Pivovaroff M.J., Kawai N.,
Tamura K., 2002, ApJ, 569, 878
\bibitem{} Gaensler B. M., Pivovaroff M. J., Garmire G. P., 2001, ApJ, 556,
L107
\bibitem{} Gallant Y.A., Achterberg A., 1999, MNRAS, 305, L6
\bibitem{} Gallant Y.A., Arons J., 1994, ApJ, 435, 230
\bibitem{} Gallant Y. A., Tuffs R. J., 1999, Pulsar Astronomy - 2000 and Beyond,
ASP Conference Series, Vol. 202; Proc. IAU Coll.\ 177 (San
Francisco: ASP). Eds.: M.\ Kramer, N.\ Wex, and N.\ Wielebinski,
p. 503
\bibitem{} Gallant Y. A., Tuffs R. J., 2002, in Neutron Stars in Supernova Remnants,
ASP Conference Series, Vol. 271, Eds.: P.O.\ Slane and B.M.\
Gaensler. ASP Conf.\ Series Vol.271, p.161
\bibitem{} Gallant Y.A., van der Swaluw E., Kirk J.G., Achterberg
A., 2002, in Neutron Stars in Supernova Remnants, Eds.: P.O.\
Slane and B.M.\ Gaensler. ASP Conf.\ Series Vol. 271, , p.\ 99
\bibitem{} Helfand D. J., Gotthelf E. V., Halpern J. P.,  2001, ApJ, 556, 380
\bibitem{} Hoshino M., Arons J, Gallant Y.A., Langdon A.B., 1992, ApJ, 390, 454
\bibitem{} Kennel C. F., Coroniti F. V., 1984, ApJ, 283, 694
\bibitem{} Kirk J.G., Guthman A.W., Gallant Y.A., Achterberg A.,
2000, ApJ, 542, 235
\bibitem{} Kirk J.G., Skj{\ae}raasen O., 2003, ApJ, in press
\bibitem{} Kundt W., Krotscheck E., 1980, A\&A, 83, 1
\bibitem{} Larrabee D.A., Lovelace R.V.E., Romanova M.M., 2002,
astroph/0210045
\bibitem{} Lu F.J., Wang Q.D., Aschenbach B., Durouchoux P., Song
L.M.,  2002, ApJ, 568, L49
\bibitem{} Lyubarsky Y.E.,  Kirk J.G., 2001, ApJ, 547, 437 (LK)
\bibitem{} Lyutikov M., 2002, astro-ph/0210353
\bibitem{} Michel F.C., 1971, Comments Astrophys.Space Phys., 3,
80
\bibitem{} Michel F.C., 1982, Rev.Mod.Phys., 54, 1
\bibitem{} Michel F.C., 1994, ApJ, 431, 397
\bibitem{} Pavlov G.G., Kargaltsev O.Y., Sanwal D., Garmire G.P., 2001, ApJ, 554, L189
\bibitem{} Rees M. J., Gunn, J. E.,  1974, MNRAS, 167, 1
\bibitem{} Romanova M. M., Lovelace R. V. E.,  1992, A\&A, 262, 26
\bibitem{} Shklovsky I. S., 1970, ApJ, 159, L77
\bibitem{} Usov V.V., 1975, ApSS, 32, 375
\bibitem{} Weisskopf C.\ et al., 2000, ApJ, 536, L81
\bibitem{} Zenitani S., Hoshino M., 2001, ApJ, 562, L63

\end{thebibliography}
\end{document}